\documentclass[useAMS,usenatbib,fleqn]{mnras}
\usepackage{graphicx,graphics,amsmath,amssymb}
\usepackage{url}
\usepackage{mathtools}

\title[The Galactic halo pulsar population]
{The Galactic halo pulsar population}

\author[Rajwade et al.]
{Kaustubh Rajwade$^{1,2,3}$\thanks{Email:kaustubh.rajwade@manchester.ac.uk}, Jayanth Chennamangalam$^{4}$, Duncan R. Lorimer$^{2,3}$ and
\newauthor
Aris Karastergiou$^{4,5,6}$\\
$^1$Jodrell Bank Centre for Astrophysics, University of Manchester, Oxford Road, Manchester M13 9PL, UK\\
$^2$Department of Physics and Astronomy, West Virginia University, PO Box 6315, Morgantown, WV 26506, USA\\
$^3$Center for Gravitational Waves and Cosmology, Chestnut Ridge Building, Morgantown, WV 26505, USA\\
$^4$Astrophysics, University of Oxford, Denys Wilkinson Building, Keble
    Road, Oxford OX1 3RH, UK\\
$^5$Department of Physics and Electronics, Rhodes University,
    PO Box 94, Grahamstown 6140, South Africa\\
$^6$Physics Department, University of the Western Cape,
    Cape Town 7535, South Africa
}

\begin{document}

\maketitle

\begin{abstract}
Most population studies of pulsars have hitherto focused on the disc of the Galaxy,
the Galactic centre, globular clusters, and nearby galaxies. It is expected
that pulsars, by virtue of their natal kicks, are also to be found in the Galactic halo. We investigate the possible
population of canonical (i.e.~non-recycled)
radio pulsars in the halo, estimating the number of such pulsars,
and the fraction that is detectable via single pulse and periodicity searches.
Additionally, we explore the distributions of flux densities and dispersion
measures of this population. We also consider the effects of different velocity models 
and the evolution of inclination angle and magnetic field on our results. We show that $\sim$33\% of all pulsars beaming towards
the Earth are in the halo but the fraction reduces to $\sim$1.5\% if we let the inclination angle and the magnetic field evolve as a falling exponential. Moreover, the fraction that is detectable is 
significantly limited by the sensitivity of surveys. This population would be most effectively probed by surveys using time-domain periodicity search algorithms. The current non-detections of pulsars in the halo can be explained if we assume that the inclination angle and magnetic field of pulsars evolve with time. We also highlight a possible confusion between bright pulses from halo pulsars and Fast Radio Bursts with low dispersion measures where further follow-up is warranted.
\end{abstract}


\section{Introduction}

Since their
discovery~\citep{he68}, about 2700 pulsars have been detected from large-scale surveys. This sample, however, represents only a small fraction of the total Galactic pulsar population. A number of studies over the years~\citep[see, e.g.,][]{fau2006,ka17} have shown that there are potentially 10$^4$--10$^5$ pulsars beaming
towards us in the Galaxy. In this paper, we investigate the extent of this population which 
resides in the Galactic halo. Although this population is hard to detect and
has received relatively little attention so far,
due to its implications on our understanding of other aspects of the population, in particular
the pulsar velocity spectrum, we are motivated to study it here.

Since pulsars are born in supernovae, even small asymmetries during the explosion can result in
strong natal kicks
\citep[see, e.g.,][]{shk1970}.
A study of pulsar birth velocities
by \cite{hob2005} showed that the velocities are consistent with a single Maxwellian distribution. However, a more recent study by \cite{ver2017a} finds evidence for a bimodal distribution. Regardless of the form of the distribution, it is widely accepted that a significant fraction of all pulsars will escape the
Galactic gravitational potential, and, depending on its age, may be found
far outside the Galactic disc. Bound pulsars, on the other hand, depending on
their velocities, can travel far outside the disc and cross the disc multiple
times during their lifetime, as they orbit the Galactic centre of mass. In this
paper, we define the halo population of pulsars as those found outside the disc of the Galaxy
and within the virial radius of the Galaxy, which we henceforth refer to as the halo population. Here, the
halo is defined to be the gaseous envelope -- in other words, the circumgalactic
medium (CGM) -- of the Galaxy out to the virial radius, which we assume to be
$\sim$200~kpc~\citep{shu2014}.

Early work on pulsars in the halo was done a few decades ago. High velocity neutron stars beyond the Galactic disc were invoked to explain gamma-ray bursts by many authors~\citep[see, e.g.,][]{li92}. The first population synthesis of pulsars in the halo was done by~\cite{jo95} where they focused on the dynamic evolution of pulsars in the Galactic potential. The authors found that over $\sim$65$\%$ of the pulsars were unbound by the virtue of their high birth velocities and could be found in the halo of the Galaxy though they did not report on the possibility of detecting them with radio surveys. In a paper dealing with the implications of the high birth velocities of pulsars, \cite{bai1999} suggested a population of old and dead neutron stars in the Galactic halo.~\cite{bai1999} proposed the use of pulsars in the halo to study the Galactic potential and the electron content of the CGM. Pulsars in globular clusters already provide us with sight lines up to a few kpc beyond the disc~\citep[see, e.g.,][]{ran2005}. Detectable halo
pulsars, on the other hand, have the potential to provide us with sight lines
that are many tens of kpc, probing farther out into the CGM. 
Much is unknown
about the main constituents of the CGM. O\textsc{vii} absorption line studies
have shown that ionized material permeates throughout the halo. For example,
\cite{gup2012} use the ionization fraction of O\textsc{vii} and the emission
measure of the CGM to compute the electron density of the medium to be
$\sim$2$\times10^{-4}$~cm$^{-3}$. Measuring the dispersion measure (DM) of
multiple halo pulsars would provide an alternate method to obtain an estimate of the
electron density of the CGM. Reliable distance estimates to these pulsars will certainly help in accurate measurement of the electron content in the halo. This would allow us to better determine the baryon
content of the universe, potentially helping solve the long-standing `missing baryon
problem' \citep{per1992,mcq2014.}

In this paper, using Monte Carlo (MC) simulations, we attempt to quantify the 
large population of pulsars in the halo. 
The outline of this paper is as follows: In \S\ref{sec_kinematic}, we discuss
the dynamics of pulsar motion and show that it can lead to a substantial halo
population, and discuss the properties of this population. In
\S\ref{sec_discussion}, we discuss detection considerations, before concluding
with the implications of these results.

\section{Simulations}\label{sec_kinematic}

It is well known \citep[see, e.g.,][]{ll94} that neutron stars (NSs) are born
with high velocities due to the violent nature of their formation in supernovae. 
In one study, \cite{hob2005} show that the three-dimensional velocities of Galactic radio pulsars follow a
Maxwell-Boltzmann distribution with a root-mean-square (rms) value of 265~km~s$^{-1}$. Later
studies~\citep{ver2017a,ver2017b} argue that the \cite{hob2005} distribution
underestimates the number of low-velocity pulsars, and describe the observed
velocities using a distribution in which 32\% of pulsars are in a Maxwellian
with a rms of 77~km~s$^{-1}$ and 68\% of pulsars are in a Maxwellian with a
rms of 320~km~s$^{-1}$. All these proposed distributions indicate that a
large fraction of pulsars have velocities greater than the local escape
velocity, i.e., they are gravitationally unbound. MC simulations by
\cite{ofe2009} show that 60--70\% of NSs
born in the Galaxy are unbound. This is in disagreement with the results of
\cite{sar2010}, who estimate that only $\sim$15--30\% of NSs are unbound, with the
disagreement attributed to different initial conditions. Depending on the
velocities of these NSs and their ages, they may be found far outside the disc
of the Galaxy. The oscillation of bound pulsars in the Galactic potential can
also take them far outside the disc, into the Galactic halo. \cite{sar2010}
estimate that $\lesssim$80\% of all NSs -- whether bound or unbound -- are to
be found in the halo.

To estimate the number of isolated radio pulsars in the halo beaming towards us, we performed MC
simulations of their evolution following the work of~\cite{fau2006}. We started
with the following simplifying assumptions: (i) A constant supernova rate of one
per century; (ii) all supernovae produce NSs; (iii) all NSs are radio
pulsars. These assumptions lead to 10$^8$ pulsars to be born in our simulations, and we evolve these pulsars over time, through the Galactic potential. We used the distributions of birth spin period and surface magnetic
field used by \cite{fau2006}, a constant pulsar radius of 10~km, and a constant
moment of inertia of 10$^{38}$~kg~m$^{2}$, to calculate the spin evolution of the
population. For the spin evolution, we considered two cases: (i) all pulsars are orthogonal rotators and the magnetic field remains constant; (ii) the magnetic field and the inclination angle evolve with time.

\begin{figure*}
\includegraphics[width=\linewidth]{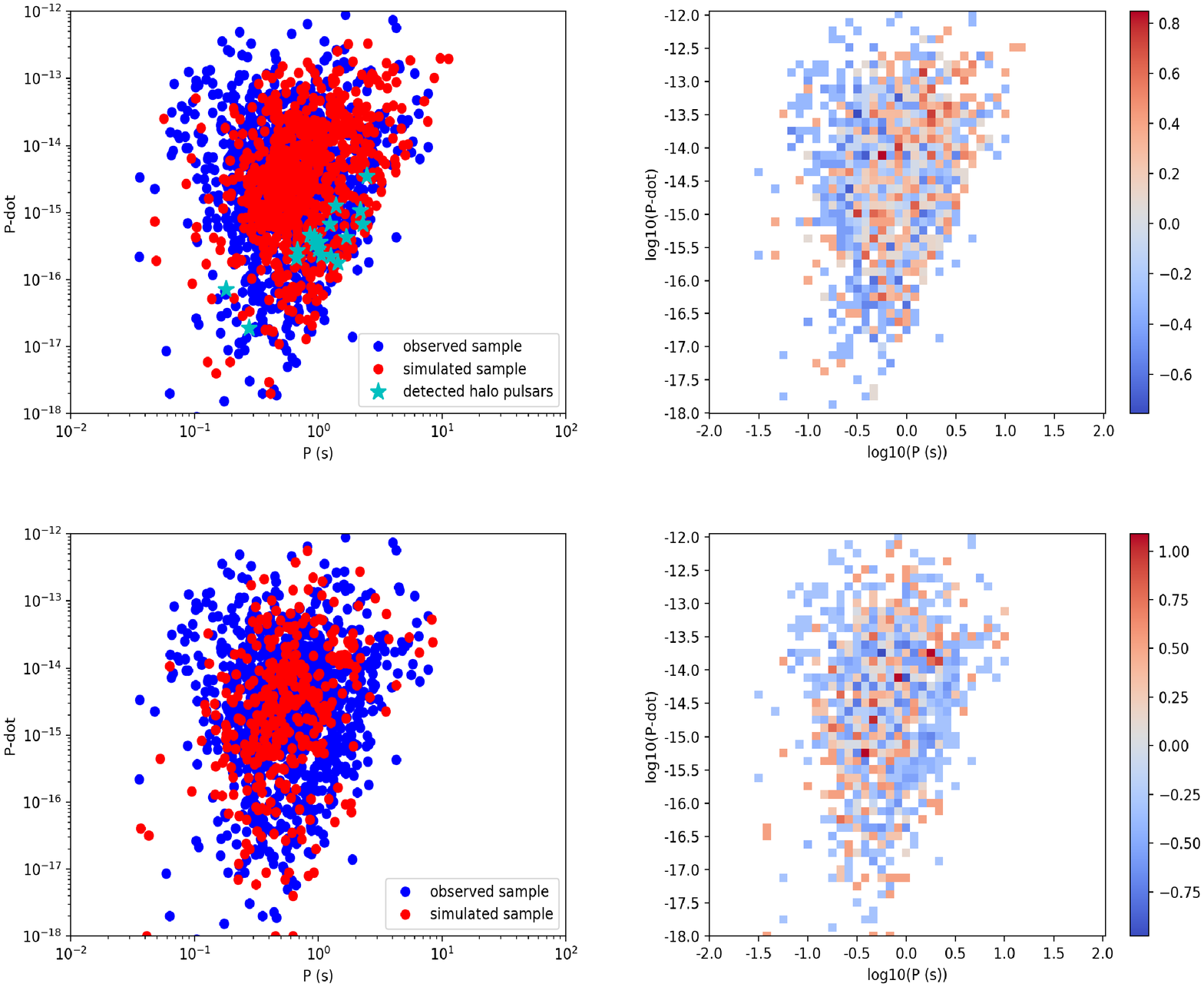}
\caption{Left column: $P-\dot{P}$ diagrams showing actual observed sample of pulsars in Parkes multi-beam survey \protect{\citep{man2001}} and detected sample in the simulations along with the detected halo pulsars in the same simulated Parkes multi-beam survey survey. Right column: The value of the goodness of fit metric \citep[Eq.~12 in][]{ka17} in $P-\dot{P}$ space. The simulations are using a unimodal birth velocity distribution. The upper row is with no evolution of the magnetic field and inclination angle and bottom row is with the evolution of magnetic field and inclination angle. The absence of halo pulsars in the lower panel is due to the lack of detections for that case (see text for details).
\label{fig:pmsurv_check}}
\end{figure*}

We then extracted those pulsars that have not crossed the death
line, i.e., radio-loud pulsars. We applied a radiation beaming correction to
extract those radio-loud pulsars beaming towards the Earth. To calculate the radio luminosity of each 
simulated pulsar, we have adopted the same scaling law as~\cite{fau2006} ($\alpha$ = -1.5 and 
$\beta$ = 0.5). To make sure that our simulations are consistent with observations, we performed a simulated Parkes multi-beam survey~\citep{man2001} on
the simulated population and compared the resulting detected sample with the observed sample of pulsars detected in the 
Parkes multi-beam survey. Figure.~\ref{fig:pmsurv_check} shows the resulting comparison. Following the methodology presented in~\cite{fau2006} and~\cite{ka17}, we have shown the goodness of fit metric \citep[Eq.~12 in][]{ka17} for both evolution scenarios. The variation of the metric is smooth and the root mean square deviations are less than 1. This suggests that our evolved population is consistent with observations. However, we note that since we are only interested in calculating the relative fractions of pulsars present in the halo and the fraction detectable, we do not focus on their actual numbers and thus, do not comment on the constraint on the total population of pulsars. For the purposes of our simulations, we have adopted a nominal birth rate of 1 pulsar per century keeping in mind that the reported values in the literature are highly uncertain due largely to distance errors~\citep[see][and  references therein.]{lor2004}.
To assign birth
locations, we followed \cite{fau2006}, who use the results of \citet{yus2004}
for radial distribution in the Galaxy, and an exponential distribution
perpendicular to the plane of the Galaxy. Fitting a simple linear model to the
Galactocentric radial escape velocity estimates of \cite{pif2014} (their
Fig.~12), we determined the fraction of unbound pulsars beaming towards the
Earth.

Given that even bound pulsars can travel outside the disc of the Galaxy and into the
halo during the course of their oscillation in the Galactic potential, and
unbound pulsars -- provided they are young -- can be found to be still in the
disc, we used the YMW16 model of the Galaxy~\citep{ymw16} to extract those
pulsars that are in the halo. Even though YMW16 is strictly a model of the
electron density in the Galaxy, for the purpose of this task, we assumed that
the model defines the \emph{extent} of the Galaxy. For each pulsar in our
sample, we computed the line-of-sight distance to the `edge' of the Galaxy as
per the model -- i.e., the distance beyond which the DM does not increase. Any
pulsar with a distance greater than this was considered to be a halo pulsar. 

The results of the MC simulations are shown in Table~\ref{tab:vmodel}.
As can be seen, the different velocity models can change the number of unbound pulsars beaming towards us by a factor of $\sim$1.5. In our MC
simulations, we have not directly computed the fraction of \emph{all} pulsars
that are unbound. However, if we assume that the population of pulsars beaming towards us is
dynamically representative of all pulsars
that are born in the Galaxy, the fraction of all pulsars that is unbound is
$\sim$9--15\%. This is consistent with the results of \cite{sar2010} who claim that $\sim$10--15\% of the neutron stars found in our Galaxy are unbound. The change in unbound fraction due to different velocity models is expected, given the change in the initial kick conditions of the young neutron star at birth. We
have computed that $\sim$30--33\% of all pulsars beaming towards the Earth are to be
found in the halo if we assume no evolution of inclination angle and magnetic field. If the inclination angle and magnetic field evolve, the fraction drops drastically to $\sim$1--3\%. The reason for such a change is that since most pulsars in the halo are old, under the evolutionary assumption, most of them are aligned rotators, causing a significant reduction in the beaming fraction. Though the evolution of magnetic field and inclination angle reduces the number of pulsars beaming towards us but does not affect their luminosity and flux distributions. The luminosity and flux distributions of halo pulsars are shown in Figure~\ref{fig_lumfluxdist} along with distributions of all pulsars beaming towards us for comparison. One can clearly see that the mean flux density of pulsars in the halo is lower than the mean flux density of all pulsars by 2 to 3 orders of magnitude, highlighting the difficulty of detecting these sources with current pulsar surveys.

\begin{figure*}
\includegraphics[width=\linewidth]{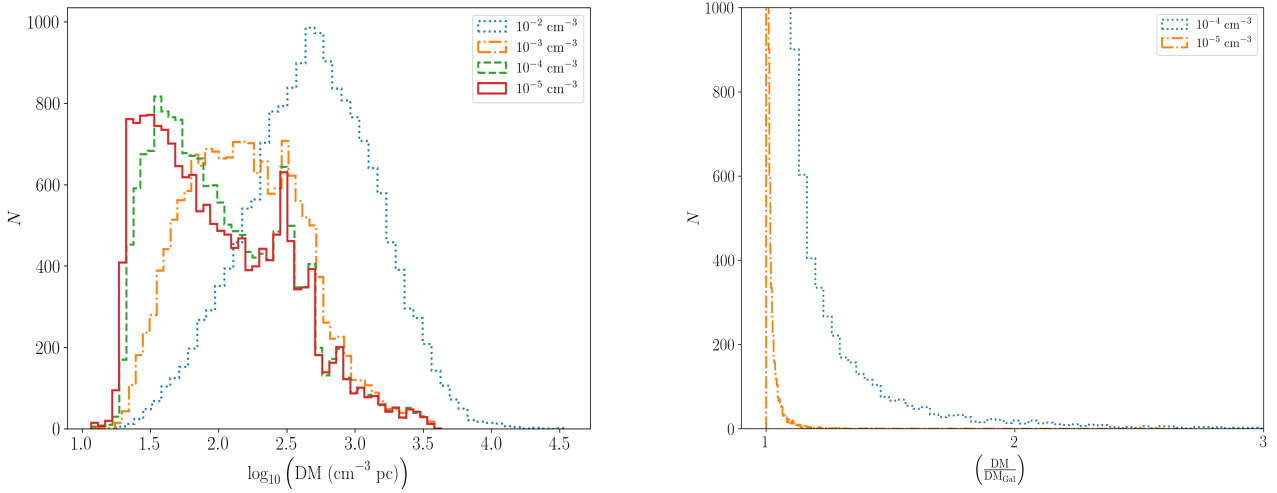}
\caption{\textbf{Left Panel}:The DM distribution of halo pulsars for various values of CGM electron
density assuming unimodal velocity distribution and no evolution of magnetic field and the inclination angle. The maximum galactic contribution is calculated using YMW16 model. For each value of electron density, we have assumed a uniform
distribution of electrons in the CGM. \textbf{Right Panel}: The histogram of halo pulsars for excess DM above the maximum Galactic contribution along that line of sight using the YMW16 model. We show histograms for two assumed electron densities in the halo. The histograms have been zoomed in to show that there are pulsars in the halo with a DM excess of a factor of two for electron density of $\sim$10$^{-4}$~cm$^{-3}$~\protect\citep{gup2012}.
\label{fig_dm_dist}}
\end{figure*}

Next, we examined the DM distribution of halo pulsars. The YMW16 model gives the DM
contribution of the Galaxy for a given line of sight. It is unclear what the
contribution of the CGM would be, for halo lines of sight, with estimates of the CGM
electron density varying from $\sim$2$\times10^{-4}$~cm$^{-3}$~\citep{gup2012} to
10$^{-2}$~\citep{shu2014}. Since there is a large uncertainty in the measured value
of electron density, we compute DM distributions for a range of electron density
values, and these are shown in Figure~\ref{fig_dm_dist}. Depending on the electron density in the halo, the figure shows that  pulsars in the halo can possess high DMs that are comparable to the DMs of pulsars in the Galactic plane and the Galactic Centre magnetar PSR J1745--2900~\citep{sha2013}. 
\begin{table*}
    \centering
    \begin{tabular}{l|r|r|r|r|r|r|}
        Velocity Distribution & Inclination Evolution & $N$ & $f_{\rm u}$ &
        $N_{\rm h}$ & $f_{\rm h,u}$ & $f_{\rm u,h}$\\
        \hline
        Unimodal Maxwellian & No &50316 &0.09 & 16898 & 0.18 & 0.65\\
        Unimodal Maxwellian & Yes &13382 &0.09 &97 &0.20 & 0.02\\
        Bimodal Maxwellian & No & 50210 & 0.15 &15126 &0.33 & 0.67\\
        Bimodal Maxwellian & Yes & 13528 & 0.15 & 174 & 0.43 & 0.04\\
        \hline
    \end{tabular}
    \caption{Representative results from a single realization of the MC
    simulation. `Unimodal Maxwellian' refers to the velocity distribution of~\citet{hob2005}, 
    while `Bimodal Maxwellian' refers to that of~\citet{ver2017a}. $N$ is the total number of potentially observable
    pulsars, $f_{\rm u}$ is the fraction of these pulsars that is unbound,
    $N_{\rm h}$ is the number of potentially observable pulsars in the
    halo, $f_{\rm h,u}$ is the fraction of these pulsars that is unbound,
    and $f_{\rm u,h}$ is the fraction of all unbound pulsars that is in the halo.
    \label{tab:vmodel}}
\end{table*}


\begin{figure}
\includegraphics[width=\linewidth]{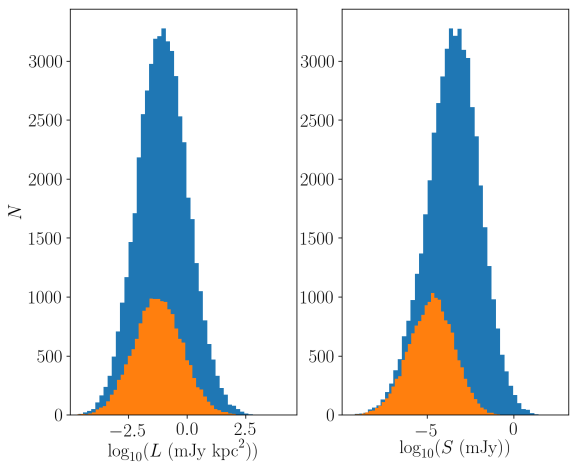}
\caption{The luminosity and flux distributions of all pulsars (blue) and halo pulsars (orange) in a single
realization of the MC simulation assuming unimodal velocity distribution and no inclination evolution.
\label{fig_lumfluxdist}}
\end{figure}

\section{Detecting Halo Pulsars}\label{sec_discussion}

Our dynamical simulations show that a significant population of all pulsars beaming towards the Earth is found in the halo. In this section,
we compute the detection statistics of this population using different search
techniques. 

\subsection{Single pulse searches}

We find the peak flux density for each of our simulated halo pulsars  as follows.
Using the luminosity scaling law given in~\cite{fau2006}, we computed the 1.4~GHz
pseudo-luminosity. Then, the flux density $S = {L}/{D^2}$,
where $L$ is the pseudo-luminosity and $D$ is the distance to the pulsar. Assuming
rectangular pulses with a fixed duty cycle $\delta = 1\%$, we computed the
average peak flux
$S_{\rm peak} = {S}/{\delta}$.
Next, following \cite{bur12}, we adopted a log-normal distribution for the peak fluxes to
sample single pulses from a given pulsar. For each pulsar, we drew $N$ peak flux
values from a log-normal flux distribution with mean equal to $S_{\rm peak}$ of
that pulsar and standard deviation equal to 10\% of $S_{\rm peak}$. Here $N$ is computed based on the observation
duration and the period of the selected pulsar. Our simulated survey observed the
entire sky visible to the telescope.
We then ran a simulated single pulse search for different surveys (see
Table~\ref{tab:detections}) on these $N$ pulses. We computed single pulse
sensitivity for each survey using Eq.~15 of \cite{cor2003}. The search was
performed on all halo pulsars beaming towards us, and those pulsars emitting detectable pulses above the
sensitivity limit of the survey were extracted. The resulting detections are reported
in Table~\ref{tab:detections}. 


We observed  that the detected pulses 
originate
only from the brightest pulsars in the halo, leading us to conclude that single
pulse searches would prove ineffective in revealing the pulsar population of the halo. The reason for this is that the distribution of the mean flux densities
of the halo pulsar population (Fig.~\ref{fig_lumfluxdist}) lies below the detection
threshold of the surveys we have considered, hence, even the brightest pulses from
these pulsars tend to not cross the threshold.

\subsection{Periodicity searches}

\begin{figure}
\includegraphics[width=\linewidth]{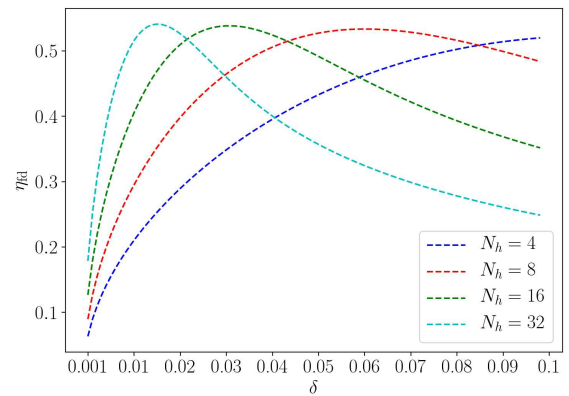}
\caption{Efficiency of Fourier domain searches compared to time domain searches for different number of harmonics to be summed as a function of pulsar duty cycle.
\label{fig:fdeff}}
\end{figure}

Next, we looked at the potential yields from periodicity searches for halo pulsars.
To simulate a frequency-domain periodicity search, we set up a search similar to the one reported
in~\cite{cor97}. For a discrete Fourier transform length $N_{\rm fft}$, flux
density $S_0$, gain $G$, system temperature $T_{\rm sys}$, sampling interval
$t_{\rm s}$, and bandwidth $\Delta \nu$, the signal-to-noise ratio 
\begin{equation}
S/N = S_0~\frac{G~\sqrt{n_{p}\Delta \nu t_{s}~N_{\rm fft}}}{ \beta T_{\rm sys}} \left(\frac{R_{l}}{\sqrt{N_{h}}}\right),
\label{eq:SNfft}
\end{equation}
where the pulsed fraction,
\begin{equation}
R_{l} = \sum_{l=1}^{N_{h}}\frac{S(l)}{S(0)},
\end{equation}
where, $S(l)$ is the amplitude of the $l^{th}$ harmonic of the pulsar and $S(0)$ is the amplitude of the DC component. Here, $\beta$ is a correction factor due to digitisation, $N_{h}$ is the number of harmonics to be summed, and $n_{\rm p}$ is the number of polarisations. We note that $R_{l}$ depends on the shape of the pulse and in our case, we assumed Gaussian pulses~\citep[for more details, see][]{cor97}. The term $ \left(\frac{R_{l}}{\sqrt{N_{h}}}\right)$ in Eq.~\ref{eq:SNfft} is equivalent to $\sqrt{\frac{1-\delta}{\delta}}$ in the time domain radiometer equation~\citep{lor2005}. Hence the efficiency between Fourier and time domain searches,
\begin{equation}
\eta_{\rm fd} = {\sqrt{\frac{1-\delta}{\delta}}} {\frac{\sqrt{N_{h}}}{R_{l}}}.
\label{eq:eff}
\end{equation}

It is important to consider the effect on the sensitivity of the search due to the incoherent summing of harmonics which results in a decrease in the $S/N$ of the detection~\citep[see][for more details]{raj2017}. Figure~\ref{fig:fdeff} shows that maximum efficiency of Fourier domain searches is about 60$\%$. This suggests that a $S/N$ threshold of 6 will result in a Fourier search that is approximately as sensitive as the time domain search. Since frequency domain searches have multiple degrees of freedom (2 times the number harmonics to be summed), the decision in the $S/N$ threshold depends on the trade-off between sensitivity and increase in the number of false positives due to Radio Frequency Interference (RFI) and the number of DM trials in the search~\citep[see][for a detailed mathematical treatment]{lor2005}. Lowering the threshold to 6 will result in a significant increase in the number of false positives rendering the search difficult to almost impossible to carry out. Hence, we decided to use a threshold of 9 for our frequency domain searches.

Using Eq.~\ref{eq:SNfft}, we computed the $S/N$ for all halo pulsars in our sample. Again, we used different surveys for simulating our periodicity searches. In our simulated search, we summed 16 harmonics as used by recent pulsar surveys. We filtered pulsars that had a $S/N$ greater than 9. 
We detected more pulsars in the periodicity searches than in the single pulse searches, with the yield increasing by a factor of $\sim$3--20, but we were only able to detect a small fraction of the total simulated pulsar population in the halo (see Table.~\ref{tab:detections}).

Recent advances in accelerated computing have brought time-domain periodicity searching back to the fore. The fast folding algorithm \citep[FFA;][]{sta1969} has been implemented in ongoing surveys~\citep{kon09,cam17}. This fact led us to examine detections based on time-domain searches for pulsars. For a given flux density, $S_{\rm 0}$, the signal to noise for time domain folding 
\begin{equation}
S/N = S_0~\frac{G~\sqrt{n_{p}\Delta \nu t_{s}~\tau_{\rm int}}}{ \beta T_{\rm sys}} \left(\sqrt{\frac{1-\delta}{\delta}}\right),
\end{equation}
where $\tau_{\rm int}$ is the integration time. We ran a time domain search for each of the halo pulsars are filtered pulsars with a $S/N$ greater than 10. Such a search gives us the best yields (see Table~\ref{tab:detections}), because we recover all the power lost in the incoherent harmonic summing in frequency-domain searches. However, we point out that the efficiency of a frequency-domain periodicity search is sensitive to the duty cycle of the pulse (see Figure.~\ref{fig:fdeff}) we do obtain the maximum efficiency for frequency domain searches as halo pulsars are expected to have small duty cycles~\citep{gil96}. Overall, our simulations show that we will need extensive time-domain searching in order to maximize our chances of finding pulsars in the halo.

\begin{table*}
    \centering
    \begin{tabular}{l|c|c|c|c|c}
        Survey & $N_{\rm sp}$ & $N_{\rm tdp}$  & $N_{\rm fdp}$ & $S_{\rm lim}$& T$_{\rm int}$ \\
         & & & & (mJy) & (s) \\
         \hline \\
        Unimodal Maxwellian birth velocity distribution \\
        \hline \\
        (i) Arecibo & 9 (0.05) &  91 (0.50)  &  28 (0.20)  & 8.1 & 600 \\
        (ii) Green Bank & 2 (0.01)  &  63 (0.40) &  25 (0.10)  & 25.1 & 600 \\
        (iii) MeerKAT & 7 (0.04)  &  57 (0.30) &  28 (0.20)  & 19.5 & 600 \\
        (iv) Parkes & 0 (0.0)  &  8 (0.05)  &  5 (0.03)  & 187.1 & 2100 \\
        (v) HTRU-N & 0 (0.0)  &  32 (0.20)  &  12 (0.07)  & 87.3 & 1500\\
        \hline \\
        Bimodal Maxwellian birth velocity distribution \\
        \hline
        (i) Arecibo & 9(0.06) & 71(0.50) & 31(0.20) & 8.1 & 600 \\
        (ii) Green Bank & 0 (0.0) & 34 (0.20) & 15 (0.09) & 25.1 & 600 \\
        (iii) MeerKAT & 7 (0.04) & 41 (0.30) & 23 (0.10) & 19.5& 600 \\
        (iv) Parkes & 0 (0.0) & 8 (0.05) & 4 (0.03) & 187.1 & 2100 \\
        (v) HTRU-N & 0 (0.0) & 19 (0.10) & 8 (0.05) & 87.3 & 1500 \\
        \hline
    \end{tabular}
    \caption{The number of pulsars detected in single pulse searches~($N_{\rm sp}$), time-domain periodicity searches~($N_{\rm tdp}$), and frequency-domain periodicity searches ($N_{\rm fdp}$), with fractions of detected halo pulsars expressed as percentages in parenthesis, for various surveys, both hypothetical, and past/ongoing, as follows: (i) A survey at Arecibo, using the ALFA receiver; (ii) A Green Bank telescope survey using its L-band receiver; (iii) MeerKAT, with its L-band receiver; (iv) Parkes survey with setup similar to the multi-beam survey~\citep{man2001}; (v) HTRU-North~\citep{bar2013}. The penultimate column (S$_{\rm lim}$) denotes the single pulse sensitivity of these surveys for a 1\% duty cycle pulse while the last column (T$_{\rm int}$) shows the integration time per pointing for each survey. For the time and frequency-domain searches, we used a signal-to-noise ratio of 10 and 9, respectively.}
    \label{tab:detections}
\end{table*}

\section{Discussion}

We have examined the pulsar population of the Galactic halo, and have shown that $\sim$33\% of all pulsars beaming towards us reside in the halo. That number reduces to $\sim$1.5\% if we let the inclination angle and the magnetic field evolve with time. The reduction in the beaming fraction is due to most pulsars in the halo being old, and hence, aligned rotators. We also observed that 9--15$\%$ of pulsars are unbound in the Galaxy. This is in agreement with the results reported in~\cite{sar2010}, although we have only considered two velocity models, compared to five used by~\cite{sar2010}. We observe that the fraction of pulsars in the halo changes by a factor of $\sim 1.5-2$ if we use a bimodal velocity distribution. This change is expected as the final velocities and positions of pulsars depend significantly on the initial conditions during birth i.e. kick velocities. We used simulated surveys to compute the number of halo pulsars that would be detectable using multiple search techniques commonly used in past and current surveys. The main conclusion from these results is the fraction of halo pulsars that would be detectable as there are still large uncertainties on the total population of pulsars beaming towards us as well as their birth rates~\citep{lor2004}. We have shown that the most sensitive searches with current instruments will be sensitive to $\sim 0.5\%$ of the total  number of pulsars in the halo beaming towards us assuming that magnetic field and the inclination angle do not evolve with time.

The number of pulsars in the halo, beaming towards us, is affected by the evolution of the inclination angle of the magnetic axis and the magnetic field with time. If we assume that the magnetic axis tends to align with the rotation axis as pulsars age~\citep{tau2001}, a significant fraction of the pulsars in the halo would be aligned rotators. 
Moreover, the change in the magnetic field would also change the torque on the rotating neutron star, thus changing the $\dot{P}$ and the final evolved period. We implemented evolution in our simulation based on work done in~\cite{tau2001}, assuming nominal timescales for axis alignment and magnetic field evolution. Our results showed that alignment reduces the number of pulsars in the halo that are beaming towards us by two orders of magnitude. This may be able to explain non-detections of halo pulsars in previous surveys. In fact, assuming that the fraction of detectable halo pulsars from surveys is the same as that for the non-evolution case (0.5$\%$), the number of detections from each survey is consistent with zero if we let the magnetic field and the inclination angle evolve with time. Detections of pulsars in the halo from  sensitive surveys in the future would be able to answer the question of whether the evolution of the inclination axis plays an important role in the pulsar life cycle.

We used multiple search techniques to search for pulsars in our simulated population. Our analysis suggests that single pulse searches would be least effective in searching for pulsars outside the Galactic disc. The reason is two-fold: (i) The probability of a pulsar in the halo to emit a bright pulse is low, based on the ages of these pulsars~\citep{Ly17}; (ii) the large distances to these sources cause even the brightest pulses to become weak. On the other hand, we found that time and frequency-domain searches would prove more effective in detecting these faint sources although the fraction of total pulsars detected would be small. If we assume that pulsars in the halo tend to be aligned rotators, our simulation suggests that even the best search techniques currently available to us would prove ineffective.

Finally, we note that this study
raises a possible connection between bright pulses from halo pulsars to Fast Radio Bursts~\citep[FRBs;][]{lor2007,tho2013}. 
FRBs are bright radio pulses of millisecond duration. which exhibit DMs that are far in excess of the expected Galactic contribution along the line of sight. Though we do know with certainty that FRBs are extragalactic, simulations by~\cite{do2015} have shown that the Galactic contribution to the DM of the FRB has been underestimated by a factor of $\sim$2. 
We note that the authors have estimated the halo contribution to the DM by integrating out to 500~kpc and there are uncertainties associated with their estimates. However, using Figure.~2 of their paper, we estimate the DM contribution of the halo out to a radius of 200~kpc which still results in an increase in the Galactic contribution to the DM by a factor of 1.5--2. Moreover, right panel of Figure.~\ref{fig_dm_dist} shows that there is a small fraction of pulsars with a DM excess of $\sim$ 2 if we assume an electron density of the halo to be similar to the value reported in~\citep{gup2012}.

Looking at the FRB catalog~\citep{pet2015b}, there are two FRBs, namely, FRB150215 and FRB010621 where \textbf the DM contribution from the Galaxy excluding the halo is about half of the total measured DM. Hence, it is possible that a bright pulse from one of the halo pulsars can be misinterpreted as an FRB. Detailed studies of physical characteristics of these pulses will help to differentiate between an FRB and a single pulse from a halo pulsar. The detection of a periodicity or a lack thereof, or identification of the host galaxy of the FRB via multi-wavelength follow-up will clearly distinguish between the two populations thus, emphasizing the need of multi-wavelength follow-ups of sporadic, bright single pulse detections.

\section*{Acknowledgements}

We thank our anonymous referee whose comments significantly improved the paper. KR acknowledges funding from the European Research Council grant under the European Union's Horizon 2020 research and innovation programme (grant agreement No. 694745), during which part of this work was done.
JC and AK thank the Leverhulme Trust for supporting this work. JC and AK would also like to thank Philipp Podsiadlowski for useful discussions. DRL was supported by NSF RII Track I award number 1458952.

\bibliographystyle{mnras}
\bibliography{ref}
\end{document}